\documentclass[reprint,superscriptaddress,sd,amsmath,amssymb]{revtex4-1}
\usepackage{hyperref}
\usepackage{float}
\usepackage{graphicx}
\usepackage{microtype}
\usepackage{booktabs}
\usepackage[inter-unit-product={\;},output-decimal-marker={.},per-mode=symbol-or-fraction,sticky-per,exponent-product = \cdot, output-product = \cdot,separate-uncertainty=true]{siunitx}
\usepackage[titletoc]{appendix}
\newcommand{\affilWuerz}{Technische Physik, Universit\"{a}t W\"{u}rzburg, Am Hubland, 97074 W\"{u}rzburg, Germany}
\newcommand{\affilInnsbruck}{Institut f\"{u}r Experimentalphysik, Universit\"{a}t Innsbruck, Technikerstra{\ss}e 25, 6020 Innsbruck, Austria}

\begin{document}
\title{Difference-frequency generation in an AlGaAs Bragg-reflection waveguide using an on-chip electrically-pumped quantum dot laser}
\author{A. Schlager}
\email[]{alexander.schlager@uibk.ac.at}
\affiliation{\affilInnsbruck}
\author{M. Götsch}
\affiliation{\affilInnsbruck}
\author{R. J. Chapman}
\affiliation{\affilInnsbruck}
\affiliation{Quantum Photonics Laboratory and Centre for Quantum Computation and Communication Technology, School of Engineering, RMIT University, Melbourne, Victoria 3000, Australia}
\author{S. Frick}
\affiliation{\affilInnsbruck}
\author{H. Thiel}
\affiliation{\affilInnsbruck}
\author{H. Suchomel}
\affiliation{\affilWuerz}
\author{M.~Kamp}
\affiliation{\affilWuerz}
\author{S.~H\"{o}fling}
\affiliation{\affilWuerz}
\affiliation{School of Physics \& Astronomy, University of St Andrews, St Andrews KY16 9SS,~UK}
\author{C.~Schneider}
\affiliation{\affilWuerz}
\affiliation{Institute of Physics, University of Oldenburg, D-26129 Oldenburg, Germany}
\author{G.~Weihs}
\affiliation{\affilInnsbruck}

\begin{abstract}
Nonlinear frequency conversion is ubiquitous in laser engineering and quantum information technology. A long-standing goal in photonics is to integrate on-chip semiconductor laser sources with nonlinear optical components. Engineering waveguide lasers with spectra that phase-match to nonlinear processes on the same device is a formidable challenge. 
Here, we demonstrate difference-frequency generation in an AlGaAs Bragg reflection waveguide which incorporates the gain medium for the pump laser in its core. We include quantum dot layers in the AlGaAs waveguide that generate electrically driven laser light at $\sim$\SI{790}{nm}, and engineer the structure to facilitate nonlinear processes at this wavelength. We perform difference-frequency generation between \SI{1540}{nm} and \SI{1630}{nm} using the on-chip laser, which is enabled by the broad modal phase-matching of the AlGaAs waveguide, and measure normalized conversion efficiencies up to \SI{0.64(21)}{\%/W/cm\squared}. Our work demonstrates a pathway towards devices that utilize on-chip active elements and strong optical nonlinearities to enable highly integrated photonic systems-on-chip.
\end{abstract}

\date{February 07, 2022}

\maketitle

Second-order ($\chi^{(2)}$) nonlinear optical processes \cite{Boyd-2003-NonlinearOptics, Christ-2013-HighGain} have been widely utilized in quantum computation, communication and metrology \cite{Kwiat-2013-BellTest, Obrien-2007-QuantumComputing, Rudolph-2016-Optimistic, OBrien-2009-Photonic, Shapiro-2020-QuantumIllumination, 2019-Lanyon-50km, 2004-Albota-Upconversion, 2013-Liscidini-Tomography, 2020-Zhong-QuantumAdvantage}. In addition, second-harmonic, difference-frequency (DFG) and sum-frequency generation, can be used to prepare and measure ultrafast laser pulses \cite{Armstrong-1967-PulsedLaser,Moulton-1986-TiSapph}, erase spectral distinguishability between single photon emitters \cite{Weber-2019-FrequencyConv} and engineer lasers at otherwise hard-to-reach wavelengths \cite{Leindecker-2011-DFG-Laser}. Today, most $\chi^{(2)}$ photonic devices rely on crystals such as BaB$_2$O$_4$, LiNbO$_3$ or KTiOPO$_4$ that exhibit high nonlinear conversion efficiencies, but are limited in scalability as they cannot directly host active optical components such as lasers and photodetectors.

Photonic circuits in the direct band-gap semiconductor AlGaAs enable integration of active components on a platform with strong $\chi^{(2)}$ nonlinearity. Significant effort has been committed to develop miniaturized, on-chip AlGaAs devices with increasing complexity and richer applications \cite{Chang-2020-AlGaAsOnInsolator,Dietrich-2016-GaAsPhotonics,Ducci-2017-Review}. In particular, AlGaAs Bragg-reflection waveguides (BRWs) facilitate phase-matching in the telecom band and exhibit strong mode confinement, which has led to large nonlinear conversion efficiencies up to \SI{2.5e-2}{\%\per W\per cm\squared} for DFG \cite{Helmy-2012-Monolithic}. Simple waveguide circuits \cite{Ducci-2018-MMI}, entangled photon pair sources \cite{Torres-2013-PolEnt, Weihs-2017-Temporally}, single photon emitters \cite{Michler-2015-Monolithic} and on-chip lasers \cite{Helmy-2020-Laser} have been demonstrated with BRWs. However, there has been limited development of devices that exploit active components, such as lasers, and the strong second order nonlinearity at the same time \cite{Ducci-2014-Electrically, Helmy-2013-Laser}. Recent work demonstrates DFG conversion efficiency of up to \SI{169}{\percent/W/cm^2} by utilizing an internal quantum well laser \cite{Helmy-2020-DFG}.

In our approach, we demonstrate an AlGaAs BRW with an embedded Al$_{0.13}$In$_{0.47}$Ga$_{0.4}$As quantum dot (QD) laser. This material composition enables the laser to emit at a wavelength around \SI{790}{nm} at room temperature and increases the gain in comparison to conventional InGaAs QDs \cite{2007-Schlereth-DotLasers,2009-Schlereth-DotLasers}. The device is engineered such that the higher-order Bragg mode phase-matches at this wavelength with the fundamental total internal reflection (TIR) mode in the telecom band. The nonlinear waveguide comprises the active gain medium in an edge-emitting cavity, forming our electrically driven pump laser. We determine the threshold current of the internal laser and use the chip temperature to tune its spectral properties, such that it pumps type-2 nonlinear processes. By injecting a telecom signal laser into the waveguide, we generate an additional telecom idler field via DFG. We reconstruct the joint spectral intensity (JSI) by spectrally resolving the generated idler field, and extract a nonlinear conversion efficiency of up to \SI{0.64(21)}{\%/W/cm\squared}. While this is a clear improvement in comparison to previously reported BRWs without internal laser (passive structures), the device requires further engineering to match the results in Ref.~\cite{Helmy-2020-DFG}.

\begin{figure}
    \centering
    \includegraphics[width=0.9\columnwidth]{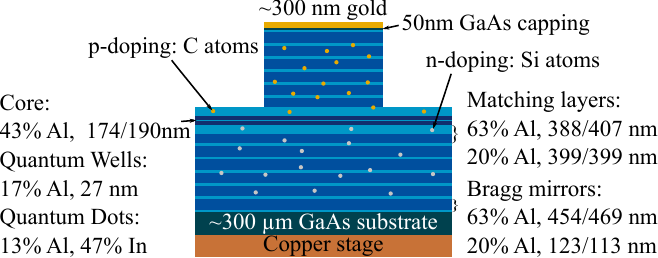}
    \caption{Schematic of the waveguide cross section. The QD layers are embedded within quantum wells with a total thickness of \SI{27}{nm}, which are centered in the core. The thicknesses have been measured via scanning electron microscopy and the layers on top of the core (first value) differ slightly from the ones underneath (second value). The structure is doped in the capping layer (\SI{20e18}{cm^{-3}}), the Bragg mirrors (\SI{2e18}{cm^{-3}}) and the matching layers (\SI{1e18}{cm^{-3}} farther from and \SI{0.5e18}{cm^{-3}} closer to the core) with C and Si atoms, respectively.}
    \label{fig:sample}
\end{figure}

We fabricate our BRWs starting with an epitaxially grown Al$_x$Ga$_{1-x}$As stack with varying aluminum concentrations $x$ and layer thicknesses \cite{Weihs-2018-Engineering}. We subsequently pattern waveguides by electron beam lithography and dry etching in Ar and Cl$_2$ plasma. In this work, we utilize a \SI{1.48}{mm} long and \SI{3.5}{\micro\metre} wide waveguide sample with an etch depth of about \SI{4.6}{\micro m}. A schematic is depicted in Fig.~\ref{fig:sample}. The waveguide geometry is designed for tight confinement of the Bragg and TIR modes and this leads to a simulated effective nonlinearity of $d_\text{eff}=\SI{56}{pm/V}$ (see Supplementary Section 1). The device is planarized with Benzocyclobuthene (BCB) to allow for depositing electrical contact pads larger than the waveguide width. The layer structure enables modal phase-matching between the TIR mode at  $\sim$\SI{1580}{nm} confined to the core and matching layers, and higher order Bragg modes at $\sim$\SI{790}{nm} confined by the Bragg-mirror stacks \cite{Helmy-2009-BRW-MatchingLayer, Weihs-2018-Engineering}. Two quantum dot layers are embedded at the center of the waveguide core and are surrounded by Al$_{0.17}$Ga$_{0.83}$As quantum wells to increase carrier accumulation. We dope the upper and lower mirror stacks as well as the substrate with carbon and silicon atoms to create a p-i-n junction. The mechanically cleaved waveguide facets have $\sim$\SI{32}{\%} reflectivity and form a cavity surrounding the gain medium. The QD dipoles are oriented to emit horizontally polarized light into the Bragg mode. 

\begin{figure}
    \centering
    \includegraphics[width=\columnwidth]{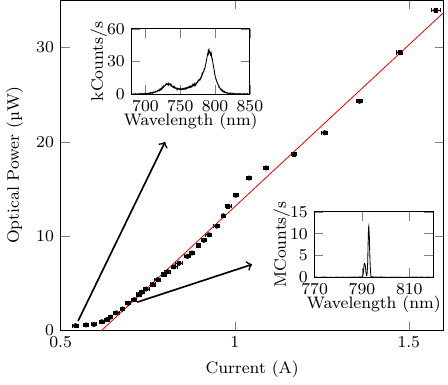}
    \caption{The optical power of the internal laser for different peak currents at \SI{3.5}{\celsius}. We use a linear fit starting with the 6th data point, which results in a threshold of \SI{618\pm4}{mA}. Deviation from the expected linear trend is likely caused by unmatched impedance between the signal generator and chip. The insets depict the measured spectra at about \SI{540}{mA} and \SI{700}{mA} respectively.}
    \label{fig:laser-combined}
\end{figure}

We use a pulsed voltage supply to characterize the QD laser with \SI{0.4}{\%} duty cycle (\SI{8}{kHz} repetition rate, \SI{500}{ns} pulse width) to reduce the thermal load on the sample while still achieving peak currents above the lasing threshold. This is necessary, as parts of the gold layer are too thin to withstand the electrical power necessary for continuous wave lasing (see Supplementary Section 2). We increase the peak voltage of the pulse generator and measure the current and generated optical power. This way, we determine the lasing threshold of \SI{618(4)}{mA}, which corresponds to a current density of \SI{11.9(5)}{kA/cm\squared}, as can be seen in Fig.~\ref{fig:laser-combined}. The insets depict the spectrum of the laser below and above the threshold current, respectively. This result of the lasing threshold is comparable to similar integrated lasers in BRWs \cite{Ducci-2014-Electrically}. However, optimized structures can offer significantly lower lasing thresholds down to a few tens of milliamperes \cite{Helmy-2020-Laser, Helmy-2020-DFG}. 

In addition to the introduction of an internal QD laser, we engineer our BRWs to enable second-order nonlinear optical processes. The frequency conversion obeys energy and momentum conservation
\begin{align}
    \label{eq:energy-conservation}
     \omega_\text{p}&=\omega_\text{s}+\omega_\text{i}\\
     \label{eq:momentum-conservation}
    k_\text{p}(\omega_\text{p})&=k_\text{s}(\omega_\text{s})+k_\text{i}(\omega_\text{i}),
\end{align}
where $\omega_j$ and $k_j(\omega_j)=n_j\omega_j/c$ are the frequency and wavenumber with $j\in\{\text{p, s, i}\}$ corresponding to pump, signal and idler and c is the speed of light. The wavenumbers are directly proportional to the effective refractive indices $n_j$ of the corresponding modes. Thus, we engineer the waveguide geometry to yield matched effective refractive indices at the desired wavelengths \cite{Helmy-2011-Advances}.

\begin{figure*}
    \centering
    \includegraphics{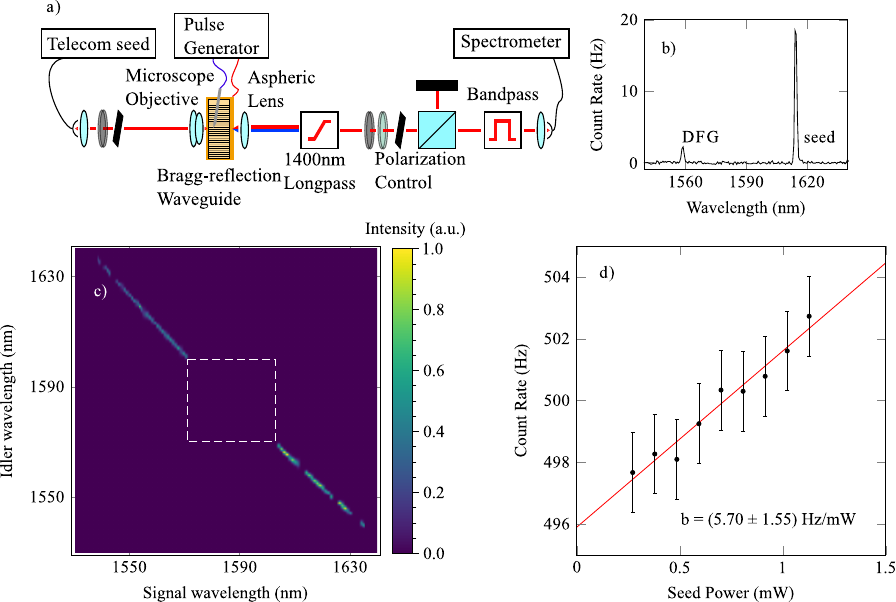}
    \caption{a) Experimental setup for performing difference frequency generation. The seed laser is coupled to the waveguide using a microscope objective. After the waveguide, we suppress the pump via a \SI{1400}{nm} longpass filter and the seed laser via polarization control (consisting of a half-wave plate, a quarter-wave plate, a polarizer and a polarizing beam splitter), as well as a \SI{40}{nm} bandpass filter centered at \SI{1550}{nm}. The remaining light is coupled into a single mode fiber and analyzed with a spectrometer and an InGaAs camera. b) Example for the measured DFG spectrum at a seed wavelength of \SI{1614}{nm} and a pump wavelength of \SI{792.9}{nm} for the quantum dot laser. Here, the seed laser is set to \SI{1614}{nm} and the corresponding DFG signal is centered around \SI{1558}{nm}. c) Joint spectral intensity reconstructed via DFG. At the regions marked with a white box, no signal could be measured due to insufficient filtering (dashed). d) Dependency of the DFG signal on the seed laser power and the corresponding slope $b$. The data represents the mean values, and the error bars the corresponding standard error of mean for an integration time of \SI{300}{s}.}
    \label{fig:dfg-combined}
\end{figure*}

We characterize the phase-matching by measuring the laser spectrum and second harmonic generation while tuning the temperature (see Supplementary Sections 3 \& 4). Due to the resulting change in the material bandgap and refractive index, the central wavelength of the laser can be shifted. This way, we determine a working point at \SI{3.5}{\celsius} and \SI{792.9}{nm} for the pump laser, at which we can investigate DFG.

We utilize the setup shown in Fig.~\ref{fig:dfg-combined}a), where we use a half-wave plate and a polarizer to control the power and polarization of the seed laser, which is coupled into the waveguide with a 100$\times$ microscope objective. The generated idler photons are then collected with a 0.68 NA aspheric lens and coupled to a spectrometer with a high-sensitivity InGaAs CCD. We use polarization, longpass, and bandpass filters to suppress the pump and seed lasers. 

We characterize the DFG process with the telecom seed laser set to \SI{1614.0}{nm} and a normalized power of \SI{909}{\micro W}. By turning on the internal QD laser as the pump, we stimulate the corresponding DFG idler emission at \SI{1558}{nm} as can be seen in Fig.~\ref{fig:dfg-combined}b). We tune the seed laser wavelength from \SIrange{1530}{1640}{nm} to measure the JSI -- the correlation of signal (seed) and idler (DFG) wavelengths -- which is enabled by the broad phase-matching of the BRW. The resulting JSI is shown in Figs.~\ref{fig:dfg-combined}c). Close to degeneracy ($\lambda_\text{s}\approx\lambda_\text{i}$), the bandpass filter can no longer suppress the seed and we therefore cannot record the DFG spectrum, as the laser oversaturates the detector. Narrow bandwidth filters at different central wavelengths would be necessary to measure the DFG in this region. Thus, we potentially miss the highest conversion efficiency, as our samples are designed for degeneracy.

Finally, we investigate the nonlinear conversion efficiency of our BRWs. We use the same settings as before, but fix the seed laser wavelength while tuning its power. This way, we measure the power dependency of the integrated DFG spectrum, as shown in Fig.~\ref{fig:dfg-combined}d). By adding a linear fit, we extract the slope $b$ in units of Hz/mW and the offset, which corresponds to the constant background (i.e. photoluminescence, dark counts and stray light). Note that we have chosen to tune the seed power since photoluminescent background light is not promoted by telecom wavelengths and is thus a constant in our measurement \cite{2021-Weihs-Noise}. We determine the conversion efficiency
\begin{equation}
    \label{eq:conv-eff}
    \eta_\text{conv}=\frac{1}{\eta}\frac{P_\text{i}}{P_\text{s}P_\text{p}L^2}=\frac{1}{\eta}\frac{b E_\gamma}{P_\text{p}L^2},
\end{equation}
where $P_{\text{p}}$, $P_{\text{s}}$ and $P_{\text{i}}$ are the normalized powers of the pump, signal and idler beams within the device (see Supplementary Section 5), respectively. The waveguide length is $L=\SI{1.48}{mm}$ and $E_\gamma$ is the energy of the frequency converted photons. We normalize the result for the total loss in our experimental setup $\eta=\SI{-56.7\pm0.7}{dB}$ for telecom light (see Supplementary Section 6). The normalized internal QD laser power is set to $P_\text{p}=\SI{2.436}{\micro W}$ and we thus extract a conversion efficiency of \SI{.64(21)}{\%/W/cm\squared}. For comparison, we perform DFG in the same waveguide using an externally coupled pump laser and present the results in Table \ref{tab:conv-eff} (more results can be found in Supplementary Section 7).

\begin{table*}
\caption{Highest measured conversion efficiencies for different measurement settings, using the external and internal laser.}
\centering
\label{tab:conv-eff}
\begin{tabular}{l c c c c c c}
\toprule
	type    & normalized power &  $\lambda_\text{p}$ &  $\lambda_\text{s}$ (nm)   &  $\lambda_\text{i}$ (nm)& b (Hz/mW) & $\eta_\text{conv}(\%/\text{W}/ \text{cm}^{2})$ \\\midrule
    internal laser  & \SI{2.436}{\micro W}   & 792.9 & 1626 & 1548  & \SI{5.70(155)}{}   & \SI{0.64(21)}{} \\
    external laser  & \SI{2.687}{mW}        & 792.9 & 1619 & 1554  & \SI{409(8)}{}      & \SI{0.041(6)}{} \\
    passive sample  & \SI{2.88}{mW}         & 767.3 & 1510 & 1560  & \SI{12.34(2)e3}{}  & \SI{0.72(12)}{}\\
\bottomrule
\end{tabular}
\end{table*}

As these results are fully normalized, we expect the same for both types of measurements with the active device. The factor of $\sim16$ difference in efficiency can be explained by the internal laser coupling more efficiently into the correct Bragg mode. Previous measurements \cite{Weihs-2015-FabryPerot}, have shown that about \SIrange{4}{8}{\percent} of the external pump power is coupled into this mode. This is a significant advantage to using the on-chip laser where the gain medium emits at an anti-node of the optical mode.

We also perform DFG using a similar BRW, but without the QD layer or doping and with comparable phase-matching wavelengths. For this, we use the external pump laser set to \SI{2.88}{mW} normalized power at \SI{767.3}{nm} and the seed laser set to \SI{1510}{nm} in the same experimental setup, which yields a conversion efficiency of \SI{0.72(12)}{\%/W/cm\squared} for a DFG emission at \SI{1560}{nm}. This value is about a factor of 18 higher than for the active sample when using the external pump. As the conversion efficiency is normalized for all losses and the modes for both devices are very similar, we explain this by a reduction of the effective nonlinearity in the active device due to its layer structure and aluminum concentrations. The type-2 process is the most efficient one in our previous structures -- typically about twice as strong as type-0 and type-1. In the active BRW, however, its signal strength has been decreased by about a factor of 10 in comparison to the other types. The cause are thickness variations of the AlGaAs layers along the length of the waveguide due to imperfections in the MBE growth process. This leads to phase-matching at only certain parts of the waveguide, thus reducing the conversion efficiency. We expect that optimization of the device \cite{Weihs-2018-Engineering} could increase the conversion efficiency by at least an order of magnitude. Furthermore, the losses within our active waveguide of \SI{-4.458}{dB/mm} for telecom light are much higher than typical values in our passive waveguides of about \SI{-2}{dB/mm}. These additional losses are caused by the QDs and doping of the sample. In addition, adding a reflective coating to the waveguide facets would improve the laser cavity and decrease the lasing threshold. Combined with thicker gold contacts, this could enable DC pumping.

In conclusion, we have demonstrated an integrated QD laser embedded in an AlGaAs BRW structure. By controlling the temperature of the chip, the internal laser wavelength can reach the nonlinear phase-matching condition, allowing for electrically pumped nonlinear processes. We have determined the nonlinear conversion efficiency of this sample via DFG, and made a comparison by externally pumping the process in this device and a passive structure. Even though our BRW does not reach a conversion efficiency as high as in devices based on a quantum well laser \cite{Helmy-2020-DFG}, a comparison with our passive structures shows the potential for improvement. In principal, QD lasers allow for a lower threshold current \cite{2011-Ellis-LowThreshold}, a higher spectral range and a large modulation bandwidth in comparison to quantum wells systems \cite{2019-Pan-QDots}. This flexibility can be important for advanced integrated photonic circuits. Furthermore, the tighter confinement of the modes in our device leads to a higher simulated effective nonlinearity of $d_\text{eff}=\SI{56}{pm/V}$. Since DFG uses the same principle and nonlinear effects as SPDC our work presented here marks another step towards fully integrated active and nonlinear photonic circuits.

\section*{Data access statement}
The data that support the findings of this study are openly available at the following DOI:
\href{https://doi.org/10.5281/zenodo.4737162}{https://doi.org/10.5281/zenodo.4737162}

\section*{Author contributions}
Conceptualization, A.S., R.J.C., S.F., G.W.; Formal analysis, A.S., M.G.; Methodology, A.S., R.J.C., S.F.; Investigation, A.S., M.G.; Resources, H.T., H.S., M.K., S.H., C.S.; Software, A.S., M.G., R.J.C.; Supervision, R.J.C., S.F., G.W.; Writing - original draft, A.S., R.J.C.; Writing - review \& editing, All Authors; Funding acquisition, C.S., G.W.;

\section*{Acknowledgments}
The authors acknowledge funding by the Austrian Science Fund (FWF) under projects Q3 (\textit{IGUANA}), project I2065 and the Special Research Program (SFB) \textit{BeyondC} project no. F7114, the DFG project no. {SCHN1376/2-1}, the EU H2020 quantum flagship program {\textit{UNIQORN} (Grant No. 820474)} and the State of Bavaria.
We thank B. Pressl for laboratory assistance. 

\section*{References}
\bibliography{dfg}

\providecommand{\noopsort}[1]{}\providecommand{\singleletter}[1]{#1}%
\begin{thebibliography}{10}
\newcommand{\enquote}[1]{``#1''}

\bibitem{Boyd-2003-NonlinearOptics}
R.~W. Boyd, \emph{Nonlinear Optics} (Academic Press, 2003).

\bibitem{Christ-2013-HighGain}
A.~Christ, B.~Brecht, W.~Mauerer, and C.~Silberhorn, \enquote{Theory of quantum
  frequency conversion and type-{II} parametric down-conversion in the
  high-gain regime,} New Journal of Physics \textbf{15}, 053038 (2013),
  \href{http://dx.doi.org/10.1088/1367-2630/15/5/053038}{doi:10.1088/1367-2630/15/5/053038}.

\bibitem{Kwiat-2013-BellTest}
B.~G. Christensen, K.~T. McCusker, J.~B. Altepeter, B.~Calkins, T.~Gerrits,
  A.~E. Lita, A.~Miller, L.~K. Shalm, Y.~Zhang, S.~W. Nam, N.~Brunner, C.~C.~W.
  Lim, N.~Gisin, and P.~G. Kwiat, \enquote{Detection-loophole-free test of
  quantum nonlocality, and applications,} Phys. Rev. Lett. \textbf{111}, 130406
  (2013),
  \href{http://dx.doi.org/10.1103/PhysRevLett.111.130406}{doi:10.1103/PhysRevLett.111.130406}.

\bibitem{Obrien-2007-QuantumComputing}
J.~L. O'Brien, \enquote{Optical quantum computing,} Science \textbf{318}, 1567
  (2007),
  \href{http://dx.doi.org/10.1126/science.1142892}{doi:10.1126/science.1142892}.

\bibitem{Rudolph-2016-Optimistic}
T.~Rudolph, \enquote{Why {I} am optimistic about the silicon-photonic route to
  quantum computing,} APL Photonics \textbf{2}, 030901 (2017),
  \href{http://dx.doi.org/10.1063/1.4976737}{doi:10.1063/1.4976737}.

\bibitem{OBrien-2009-Photonic}
J.~L. O'Brien, A.~Furusawa, and J.~Vu{\v{c}}kovi{\'{c}}, \enquote{{Photonic
  quantum technologies},} Nature Photonics \textbf{3}, 687--695 (2009),
  \href{http://dx.doi.org/10.1038/nphoton.2009.229}{doi:10.1038/nphoton.2009.229}.

\bibitem{Shapiro-2020-QuantumIllumination}
J.~H. Shapiro, \enquote{{The Quantum Illumination Story},} IEEE Aerospace and
  Electronic Systems Magazine \textbf{35}, 8--20 (2020),
  \href{http://dx.doi.org/10.1109/MAES.2019.2957870}{doi:10.1109/MAES.2019.2957870}.

\bibitem{2019-Lanyon-50km}
V.~Krutyanskiy, M.~Meraner, J.~Schupp, V.~Krcmarsky, H.~Hainzer, and B.~P.
  Lanyon, \enquote{Light-matter entanglement over 50 km of optical fibre,} npj
  Quantum Information \textbf{5}, 1--5 (2019),
  \href{http://dx.doi.org/10.1038/s41534-019-0186-3}{doi:10.1038/s41534-019-0186-3}.

\bibitem{2004-Albota-Upconversion}
M.~A. Albota and F.~N.~C. Wong, \enquote{Efficient single-photon counting at
  1.55 {\textmu}m by means of frequency upconversion,} Opt. Lett. \textbf{29},
  1449--1451 (2004),
  \href{http://dx.doi.org/10.1364/OL.29.001449}{doi:10.1364/OL.29.001449}.

\bibitem{2013-Liscidini-Tomography}
M.~Liscidini and J.~E. Sipe, \enquote{Stimulated emission tomography,} Phys.
  Rev. Lett. \textbf{111}, 193602 (2013),
  \href{http://dx.doi.org/10.1103/PhysRevLett.111.193602}{doi:10.1103/PhysRevLett.111.193602}.

\bibitem{2020-Zhong-QuantumAdvantage}
H.-S. Zhong, H.~Wang, Y.-H. Deng, M.-C. Chen, L.-C. Peng, Y.-H. Luo, J.~Qin,
  D.~Wu, X.~Ding, Y.~Hu, P.~Hu, X.-Y. Yang, W.-J. Zhang, H.~Li, Y.~Li,
  X.~Jiang, L.~Gan, G.~Yang, L.~You, Z.~Wang, L.~Li, N.-L. Liu, C.-Y. Lu, and
  J.-W. Pan, \enquote{Quantum computational advantage using photons,} Science
  \textbf{370}, 1460--1463 (2020),
  \href{http://dx.doi.org/10.1126/science.abe8770}{doi:10.1126/science.abe8770}.

\bibitem{Armstrong-1967-PulsedLaser}
J.~A. Armstrong, \enquote{{Measurement of picosecond laser pulse widths},}
  Applied Physics Letters \textbf{10}, 16--18 (1967),
  \href{http://dx.doi.org/10.1063/1.1754787}{doi:10.1063/1.1754787}.

\bibitem{Moulton-1986-TiSapph}
P.~F. Moulton, \enquote{{Spectroscopic and laser characteristics of
  Ti:Al{\_}2O{\_}3},} Journal of the Optical Society of America B \textbf{3},
  125 (1986),
  \href{http://dx.doi.org/10.1364/JOSAB.3.000125}{doi:10.1364/JOSAB.3.000125}.

\bibitem{Weber-2019-FrequencyConv}
J.~H. Weber, B.~Kambs, J.~Kettler, S.~Kern, J.~Maisch, H.~Vural, M.~Jetter,
  S.~L. Portalupi, C.~Becher, and P.~Michler, \enquote{{Two-photon interference
  in the telecom C-band after frequency conversion of photons from remote
  quantum emitters},} Nature Nanotechnology \textbf{14}, 23--26 (2019),
  \href{http://dx.doi.org/10.1038/s41565-018-0279-8}{doi:10.1038/s41565-018-0279-8}.

\bibitem{Leindecker-2011-DFG-Laser}
N.~Leindecker, A.~Marandi, R.~L. Byer, and K.~L. Vodopyanov,
  \enquote{{Broadband degenerate OPO for mid-infrared frequency comb
  generation},} Optics Express \textbf{19}, 6296 (2011),
  \href{http://dx.doi.org/10.1364/OE.19.006296}{doi:10.1364/OE.19.006296}.

\bibitem{Chang-2020-AlGaAsOnInsolator}
L.~Chang, W.~Xie, H.~Shu, Q.~F. Yang, B.~Shen, A.~Boes, J.~D. Peters, W.~Jin,
  C.~Xiang, S.~Liu, G.~Moille, S.~P. Yu, X.~Wang, K.~Srinivasan, S.~B. Papp,
  K.~Vahala, and J.~E. Bowers, \enquote{{Ultra-efficient frequency comb
  generation in AlGaAs-on-insulator microresonators},} Nature Communications
  \textbf{11}, 1--8 (2020),
  \href{http://dx.doi.org/10.1038/s41467-020-15005-5}{doi:10.1038/s41467-020-15005-5}.

\bibitem{Dietrich-2016-GaAsPhotonics}
C.~P. Dietrich, A.~Fiore, M.~G. Thompson, M.~Kamp, and S.~H\"ofling,
  \enquote{Gaas integrated quantum photonics: {Towards} compact and
  multi-functional quantum photonic integrated circuits,} Laser Photon Rev.
  \textbf{10}, 870 (2016),
  \href{http://dx.doi.org/10.1002/lpor.201500321}{doi:10.1002/lpor.201500321}.

\bibitem{Ducci-2017-Review}
A.~Orieux, M.~A. Versteegh, K.~D. J{\"{o}}ns, and S.~Ducci,
  \enquote{{Semiconductor devices for entangled photon pair generation: A
  review},} Reports on Progress in Physics \textbf{80} (2017),
  \href{http://dx.doi.org/10.1088/1361-6633/aa6955}{doi:10.1088/1361-6633/aa6955}.

\bibitem{Helmy-2012-Monolithic}
P.~Abolghasem, {Jun-Bo Han}, {Dongpeng Kang}, B.~J. Bijlani, and A.~S. Helmy,
  \enquote{{Monolithic Photonics Using Second-Order Optical Nonlinearities in
  Multilayer-Core Bragg Reflection Waveguides},} IEEE Journal of Selected
  Topics in Quantum Electronics \textbf{18}, 812--825 (2012),
  \href{http://dx.doi.org/10.1109/JSTQE.2011.2135841}{doi:10.1109/JSTQE.2011.2135841}.

\bibitem{Ducci-2018-MMI}
J.~Belhassen, F.~Baboux, Q.~Yao, M.~Amanti, I.~Favero, A.~Lema{\^{i}}tre, W.~S.
  Kolthammer, I.~A. Walmsley, and S.~Ducci, \enquote{{On-chip III-V monolithic
  integration of heralded single photon sources and beamsplitters},} Applied
  Physics Letters \textbf{112}, 071105 (2018),
  \href{http://dx.doi.org/10.1063/1.5015951}{doi:10.1063/1.5015951}.

\bibitem{Torres-2013-PolEnt}
A.~Vall\'{e}s, M.~Hendrych, J.~Svozil\'{i}k, R.~Machulka, P.~Abolghasem,
  D.~Kang, B.~J. Bijlani, A.~S. Helmy, and J.~P. Torres, \enquote{Generation of
  polarization-entangled photon pairs in a {Bragg} reflection waveguide,} Opt.
  Express \textbf{21}, 10841 (2013),
  \href{http://dx.doi.org/10.1364/OE.21.010841}{doi:10.1364/OE.21.010841}.

\bibitem{Weihs-2017-Temporally}
A.~Schlager, B.~Pressl, K.~Laiho, H.~Suchomel, M.~Kamp, S.~H{\"o}fling,
  C.~Schneider, and G.~Weihs, \enquote{Temporally versatile polarization
  entanglement from {Bragg} reflection waveguides,} Opt. Lett. \textbf{42},
  2102 (2017),
  \href{http://dx.doi.org/10.1364/OL.42.002102}{doi:10.1364/OL.42.002102}.

\bibitem{Michler-2015-Monolithic}
K.~D. Jöns, U.~Rengstl, M.~Oster, F.~Hargart, M.~Heldmaier, S.~Bounouar, S.~M.
  Ulrich, M.~Jetter, and P.~Michler, \enquote{Monolithic on-chip integration of
  semiconductor waveguides, beamsplitters and single-photon sources,} Journal
  of Physics D: Applied Physics \textbf{48}, 085101 (2015),
  \href{http://dx.doi.org/10.1088/0022-3727/48/8/085101}{doi:10.1088/0022-3727/48/8/085101}.

\bibitem{Helmy-2020-Laser}
B.~Janjua, M.~Iu, Z.~Y.~P. Charles, E.~Chen, and A.~S. Helmy,
  \enquote{{Passively mode-locked Bragg lasers with 64 GHz sub-300 fs pulses at
  785 nm},} IEEE Photonics Technology Letters \textbf{1135}, 1--1 (2020),
  \href{http://dx.doi.org/10.1109/lpt.2020.3013815}{doi:10.1109/lpt.2020.3013815}.

\bibitem{Ducci-2014-Electrically}
F.~Boitier, A.~Orieux, C.~Autebert, A.~Lema{\^\i}tre, E.~Galopin, C.~Manquest,
  C.~Sirtori, I.~Favero, G.~Leo, and S.~Ducci, \enquote{Electrically injected
  photon-pair source at room temperature,} Phys. Rev. Lett. \textbf{112},
  183901 (2014),
  \href{http://dx.doi.org/10.1103/PhysRevLett.112.183901}{doi:10.1103/PhysRevLett.112.183901}.

\bibitem{Helmy-2013-Laser}
B.~J. Bijlani, P.~Abolghasem, and A.~S. Helmy, \enquote{Semiconductor optical
  parametric generators in isotropic semiconductor diode lasers,} Appl. Phys.
  Lett. \textbf{103}, 091103 (2013),
  \href{http://dx.doi.org/10.1063/1.4819736}{doi:10.1063/1.4819736}.

\bibitem{Helmy-2020-DFG}
M.~L. Iu, N.~Zareian, D.~Kang, E.~Chen, P.~Charles, B.~Janjua, Y.~Akasaka,
  T.~Ikeuchi, and A.~S. Helmy, \enquote{Electrically pumped efficient broadband
  cw frequency conversion in diode lasers using bulk $\chi^2$,} APL Photonics
  \textbf{5}, 011301 (2020),
  \href{http://dx.doi.org/10.1063/1.5122968}{doi:10.1063/1.5122968}.

\bibitem{2007-Schlereth-DotLasers}
T.~W. Schlereth, C.~Schneider, W.~Kaiser, S.~Höfling, and A.~Forchel,
  \enquote{Low threshold, high gain algainas quantum dot lasers,} Applied
  Physics Letters \textbf{90}, 221113 (2007),
  \href{http://dx.doi.org/10.1063/1.2745200}{doi:10.1063/1.2745200}.

\bibitem{2009-Schlereth-DotLasers}
T.~W. {Schlereth}, C.~{Schneider}, S.~{Gerhard}, S.~{Hofling}, and
  A.~{Forchel}, \enquote{Short-wavelength (760–920 nm) algainas quantum dot
  lasers,} IEEE Journal of Selected Topics in Quantum Electronics \textbf{15},
  792--798 (2009),
  \href{http://dx.doi.org/10.1109/JSTQE.2008.2011493}{doi:10.1109/JSTQE.2008.2011493}.

\bibitem{Weihs-2018-Engineering}
B.~Pressl, K.~Laiho, H.~Chen, T.~G{\"{u}}nthner, A.~Schlager, S.~Auchter,
  H.~Suchomel, M.~Kamp, S.~H{\"{o}}fling, C.~Schneider, and G.~Weihs,
  \enquote{{Semi-automatic engineering and tailoring of high-efficiency
  Bragg-reflection waveguide samples for quantum photonic applications},}
  Quantum Science and Technology \textbf{3}, 024002 (2018),
  \href{http://dx.doi.org/10.1088/2058-9565/aaa2a2}{doi:10.1088/2058-9565/aaa2a2}.

\bibitem{Helmy-2009-BRW-MatchingLayer}
P.~Abolghasem and A.~S. Helmy, \enquote{{Matching Layers in Bragg Reflection
  Waveguides for Enhanced Nonlinear Interaction},} {IEEE} Journal of Quantum
  Electronics \textbf{45}, 646--653 (2009),
  \href{http://dx.doi.org/10.1109/jqe.2009.2013118}{doi:10.1109/jqe.2009.2013118}.

\bibitem{Helmy-2011-Advances}
A.~S. Helmy, P.~Abolghasem, J.~Stewart~Aitchison, B.~J. Bijlani, J.~Han, B.~M.
  Holmes, D.~C. Hutchings, U.~Younis, and S.~J. Wagner, \enquote{Recent
  advances in phase matching of second-order nonlinearities in monolithic
  semiconductor waveguides,} Laser Photonics Rev. \textbf{5}, 272 (2011),
  \href{http://dx.doi.org/10.1002/lpor.201000008}{doi:10.1002/lpor.201000008}.

\bibitem{2021-Weihs-Noise}
S.~Auchter, A.~Schlager, H.~Thiel, K.~Laiho, B.~Pressl, H.~Suchomel, M.~Kamp,
  S.~Hoefling, C.~Schneider, and G.~Weihs, \enquote{Understanding
  photoluminescence in semiconductor bragg-reflection waveguides,} Journal of
  Optics \textbf{23}, 035801 (2021),
  \href{http://dx.doi.org/10.1088/2040-8986/abd888}{doi:10.1088/2040-8986/abd888}.

\bibitem{Weihs-2015-FabryPerot}
B.~Pressl, T.~G\"{u}nthner, K.~Laiho, J.~Ge{\ss}ler, M.~Kamp, S.~H\"{o}fling,
  C.~Schneider, and G.~Weihs, \enquote{Mode-resolved {Fabry-Perot} experiment
  in low-loss {Bragg-reflection} waveguides,} Opt. Express \textbf{23}, 33608
  (2015),
  \href{http://dx.doi.org/10.1364/OE.23.033608}{doi:10.1364/OE.23.033608}.

\bibitem{2011-Ellis-LowThreshold}
B.~Ellis, M.~A. Mayer, G.~Shambat, T.~Sarmiento, J.~Harris, E.~E. Haller, and
  J.~Vu{\v{c}}kovi{\'c}, \enquote{Ultralow-threshold electrically pumped
  quantum-dot photonic-crystal nanocavity laser,} Nature photonics \textbf{5},
  297--300 (2011).

\bibitem{2019-Pan-QDots}
S.~Pan, V.~Cao, M.~Liao, Y.~Lu, Z.~Liu, M.~Tang, S.~Chen, A.~Seeds, and H.~Liu,
  \enquote{Recent progress in epitaxial growth of {III}{\textendash}v
  quantum-dot lasers on silicon substrate,} Journal of Semiconductors
  \textbf{40}, 101302 (2019),
  \href{http://dx.doi.org/10.1088/1674-4926/40/10/101302}{doi:10.1088/1674-4926/40/10/101302}.

\end{thebibliography}

\end{document}